\documentstyle[amssymb,aps,floats,psfig,epsf,twocolumn]{revtex}
\begin{document}
\draft
\title{Quasiparticle anisotropy and pseudogap formation
from the weak-coupling renormalization group point of view}
\author{A. A. Katanin$^{a,b}$ and A. P. Kampf$^a$}
\address{\sl
$^a$ Institut f\"ur Physik, Theoretische Physik III,
Elektronische Korrelationen und Magnetismus,\\
Universit\"at Augsburg, 86135 Augsburg, Germany\\
$^b$ Institute of Metal Physics, 620219 Ekaterinburg, Russia}
\address{~
\parbox{14cm}{\rm
\medskip
\vskip0.2cm
Using the one-loop functional renormalization group technique we evaluate the
self-energy in the weak-coupling regime of the 2D $t$-$t^{\prime}$ Hubbard
model. At van Hove (vH) band fillings and at low temperatures the quasiparticle
weight along the Fermi surface (FS) continuously vanishes on approaching the
($\pi ,0$) point where the quasiparticle concept is invalid. Away from vH
band fillings the quasiparticle peak is formed inside an anisotropic
pseudogap and the self-energy has the conventional Fermi-liquid
characteristics near the Fermi level. The spectral weight of the
quasiparticle features is reduced on parts of the FS between the near vicinity
of hot spots and the FS points closest to $(\pi,0)$ and $(0,\pi)$.
 }}

\maketitle

\tighten

The apparent non-Fermi liquid (nFL) normal-state features and the intriguing
pseudogap structures remain a key issue in the ongoing research activities
on high-T$_c$ superconductors. Angle resolved photoemission spectroscopy
(ARPES) keeps providing new and increasingly detailed insight into the
nature of the correlated electronic states responsible for the
unconventional normal state properties and the superconducting pairing state%
\cite{ARPES,ARPESAD}. Refined ARPES data continue to serve as challenging
test grounds for theoretical concepts developed for studying
correlated lattice electrons in two dimensions (2D).

The pseudogap phenomenon in cuprates manifests itself by the absence of
well-defined quasiparticles (qps) near the ($\pi ,0$) point of the Brillouin
zone (BZ) and the seeming partial disappearance of the FS
\cite{ARPESAD}. One direction of recent research has been to seek the origin
of these apparent non-FL characteristics and pseudogap structures in the
competing nature of different ordering tendencies indicative for the
existence of quantum critical points in the ground-state phase diagram \cite
{Varma,Loram}.

On the other hand, the vicinity to the transition from a metal to an
antiferromagnetic (AF) insulator in doped cuprates demands the understanding
of the role of AF spin fluctuations on the evolution of spectral properties
on the metallic side of the metal-insulator transition. Anisotropic
spectral features naturally arise within the 2D Hubbard model near its AF
instability even in the weak-coupling regime \cite{Kampf}; they were studied
within the spin-fermion model \cite{Pines}, fluctuation exchange
\cite{FLEX,FLEX1} and two-particle self-consistent (TPSC) approximations
\cite{TPSC}
and more recently cluster extensions of the dynamical mean-field theory (DMFT)
\cite{DCA} and cluster perturbation theory \cite{Senechal}. QMC studies on
finite clusters\cite{White} provided valuable insight into the doping
evolution of the spectral function on a sparse momentum grid in the strong to
intermediate coupling regime, but they were not able to trace the pseudogap
formation into the weak-coupling regime. The results of previous studies agree
with each other only for the case of perfect nesting (next-nearest
neighbor hopping $t^{\prime }=0$ and half-filling) where a pseudogap
opens everywhere around the FS upon cooling. The experimentally most
relevant case of a non-nested FS remains an issue of debate.

From the apparent necessity for the development of alternative methods for the
analysis of correlated electron systems in 2D, functional renormalization
group (fRG) techniques have proven to be most promising additional tools
with an emphasis on the identification of the leading electronic
instabilities in the weak-coupling regime of the 2D Hubbard model and its
extensions \cite{Zanchi,Metzner,SalmHon,SalmHon1,UVJ}.
Although the electronic on-shell lifetime, the FS shift and the approximate
qp weight calculated from the imaginary part of the self-energy at the lowest
Matsubara frequency were recently considered within the many-patch fRG
analysis\cite{SalmHon2}, these quantities alone are only indicative but
not sufficient to trace the evolution of the spectral properties in
the vicinity of the AF instability.

In this letter we use the one-loop fRG scheme for the $t$-$t^{\prime}$
weak-coupling Hubbard model to evaluate the self-energy on the real frequency
axis. We show that at vH band fillings the qp weight along the FS continuously
vanishes from the BZ diagonal towards the ($\pi ,0$) point where the qp
concept is invalid. The qp weight suppression is accompanied by the growth
of two additional incoherent peaks in the spectral function, from which an
anisotropic pseudogap originates. On moving away from vH band fillings, a
qp peak with small weight is formed inside the pseudogap. This peak regains
spectral weight from the incoherent pseudogap features on moving towards the BZ
diagonal. These results offer a complementary weak coupling view of pseudogap
formation from the RG perspective with new implications for the doped
AF-insulator regime of the Hubbard model.

Specifically, we consider the Hubbard model for $N_e$ electrons on a square
lattice
\begin{equation}
H=-\sum_{ij\sigma }t_{ij}c_{i\sigma }^{\dagger }c_{j\sigma
}+U\sum_in_{i\uparrow }n_{i\downarrow }-(\mu -4t^{\prime })N_e  \label{H}
\end{equation}
where the hopping amplitude $t_{ij}=t$ for nearest neighbor (nn) sites $i$
and $j$ and $t_{ij}=-t^{\prime }$ for next-nn sites ($t,t^{\prime }>0$); for
convenience we have shifted the chemical potential $\mu $ by $4t^{\prime }$.
We apply the fRG approach for one-particle-irreducible functions with a
sharp momentum cutoff\cite{SalmHon}. This technique considers the effective
action obtained by integrating out modes with energy $|\varepsilon _{{\bf k}%
}|\geq \Lambda $ where $\varepsilon _{{\bf k}}=-2t(\cos k_x+\cos
k_y)+4t^{\prime }(\cos k_x\cos k_y+1)-\mu $ is the electronic dispersion; $%
\Lambda $ ($0<\Lambda <\Lambda _0=\max |\varepsilon _{{\bf k}}|$) is the
cutoff parameter. This procedure is used in the weak-coupling regime for
small and intermediate $t^{\prime }\lesssim 0.3t$ when the ferromagnetic
instability is absent \cite{SalmHon1}. In this scheme the flow of the
self-energy $\Sigma _\Lambda ({\bf k},i\omega )$ is given by
\begin{equation}
\frac{d\Sigma _\Lambda }{d\Lambda }=V_\Lambda \circ S_\Lambda ,
\label{OneLoopSE}
\end{equation}
where $\circ $ is a short notation for the summation over momentum-,
frequency- and spin-variables, see e.g. Ref. \cite{SalmHon}. The
renormalization of the electron-electron interaction vertex $V_\Lambda $ at
one-loop order is given by
\begin{equation}
\frac{dV_\Lambda }{d\Lambda }=V_\Lambda \circ (G_\Lambda \circ S_\Lambda
+S_\Lambda \circ G_\Lambda )\circ V_\Lambda .  \label{OneLoop}
\end{equation}
The propagators $G_\Lambda $ and $S_\Lambda \ $are defined by
\begin{equation}
\left\{
\begin{array}{c}
G_\Lambda  \\
S_\Lambda
\end{array}
\right\} ({\bf k},i\omega _n)=\left\{
\begin{array}{c}
\theta (|\varepsilon _{{\bf k}}|-\Lambda ) \\
-\delta (|\varepsilon _{{\bf k}}|-\Lambda )
\end{array}
\right\} \frac 1{i\omega _n-\varepsilon _{{\bf k}}}.  \label{GS}
\end{equation}

\begin{figure}[t!]
\psfig{file=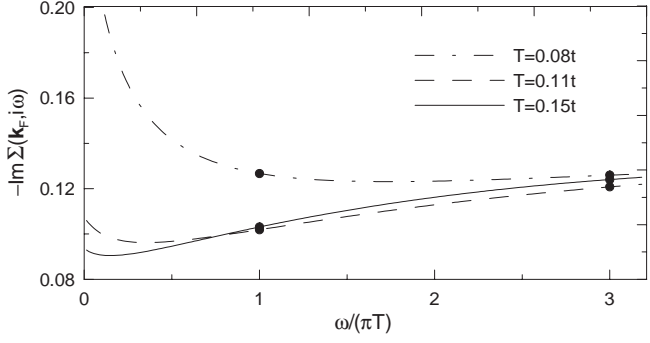,width=90mm,silent=} \vspace{2mm}
\caption{Imaginary part of the self-energy at imaginary frequencies within
the one-loop RG approach at $U=2t$, $t^{\prime}/t=0.1$ and vH band filling
n=0.92. ${\bf k}_F$ is chosen in the patch closest to the vH point at $(\pi
,0)$. The dots show the positions of the first and second Matsubara
frequency.}
\label{fig:Fig1}
\end{figure}

We neglect the influence of the self-energy and the FS shift on the RG flow
and therefore the self-energy is not included in the Green functions (\ref
{GS}). Eqs. (\ref{OneLoopSE}) and (\ref{OneLoop})\ have to be solved with
the initial conditions $V_{\Lambda _0}=U$ and $\Sigma _{\Lambda _0}=0.$
Since also the frequency dependence of the vertices is neglected, it is
convenient to reinsert, following Ref. \cite{SalmHon2}, the vertex from Eq. (%
\ref{OneLoop}) into Eq. (\ref{OneLoopSE}) to obtain

\begin{equation}
\frac{d\Sigma _\Lambda }{d\Lambda }=S_\Lambda \circ \int\limits_\Lambda
^{\Lambda _0}d\Lambda ^{\prime }[V_{\Lambda ^{\prime }}\circ (G_{\Lambda
^{\prime }}\circ S_{\Lambda ^{\prime }}+S_{\Lambda ^{\prime }}\circ
G_{\Lambda ^{\prime }})\circ V_{\Lambda ^{\prime }}].  \label{Se2}
\end{equation}

\noindent To solve Eqs. (\ref{OneLoop}) and (\ref{Se2}) numerically we divide
the momentum space into 48 patches with the same patching scheme as in Refs.
\cite{SalmHon,SalmHon1}. The self-energy on the real axis is obtained by
analytical continuation using Pad\'e approximants \cite{Pade}. To resolve
fine structures close to the Fermi level, we take advantage of Eq. (\ref{Se2}%
) that for frequency-independent vertices the self-energy can be calculated
at arbitrary frequencies on the imaginary axis. A sample result for ${\bf k}%
_F$ in the first FS patch closest to the ($\pi ,0$) point is shown in Fig.
1. Obviously, the imaginary part of the self-energy behaves
non-monotonically at frequencies $\omega<\omega_1=\pi T$ with $[\partial$ Im$%
\Sigma({\bf k}_F,i \omega)/\partial\omega]_{\omega\rightarrow0^{+}}>0$. By
the Cauchy-Riemann conditions for an analytic function this necessarily
implies $[\partial $Re$\Sigma ({\bf k}_F,\omega )/\partial \omega ]_{\omega
\rightarrow i0^{+}}>0$ and thus the breakdown of the qp concept at this
particular value of ${\bf k}_F$. Evaluating the self-energy at the Matsubara
frequencies $i \omega_n$ only (circles in Fig. 1) this behavior is missed.
Therefore, for the analytical continuation we use a dense set of frequencies
on the imaginary axis at small $|\omega |\lesssim t$ and a set of $i\omega
_n $ for $|\omega |\gg t.$ This procedure can be viewed as an analytical
continuation first from Matsubara frequencies $i\omega _n$ to some suitably
chosen set of points on the imaginary axis and a subsequent continuation to
the real axis using Pad\'e approximants. The quality of the approximants was
checked by both the analysis of the analyticity in the upper half-plane \cite
{Pade} and by requiring the fulfillment of the sum rule for the resulting
Green function.

\begin{figure}[t!]
\psfig{file=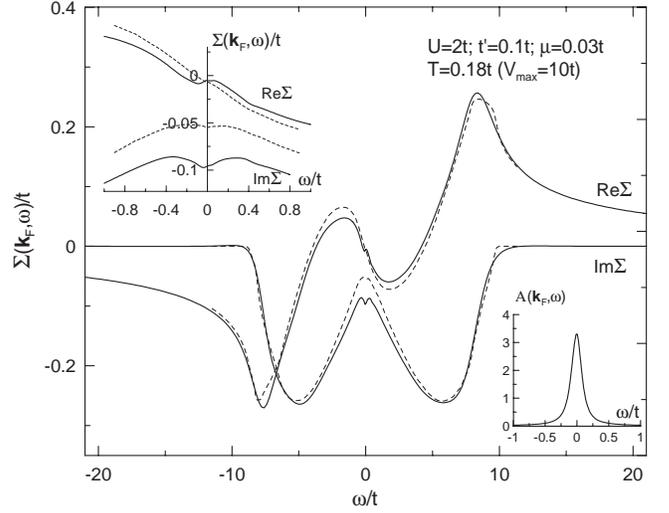,width=90mm,silent=} \vspace{2mm}
\caption{Self-energy in SOPT (dashed line) and one-loop RG (solid line) at $%
U=2t$, $t^{\prime }/t=0.1$, and vH band filling n=0.92 at $T=0.17t$ ($%
V_{max}=10t$). ${\bf k}_F^1=(3.100,0.003)$ is chosen in the first patch,
closest to the vHs. The insets show the self-energy and the spectral
function at small frequencies.}
\label{fig:Fig2}
\end{figure}

First we consider the results at the vH band filling ($\mu =0$) for $%
t^{\prime }=0.1t$ and $U=2t$. The self-energy at temperature $T=0.17t$
calculated within fRG together with the result in second-order perturbation
theory (SOPT, which is obtained by the replacement $V\rightarrow U$ in Eq. (%
\ref{Se2})) is shown in Fig. 2. Remarkably, in the first patch closest to
the vH singularity (vHs) at ($\pi ,0$), SOPT and fRG results show both a
minimum of Im$\Sigma ({\bf k}_F,\omega )$ at the Fermi level $\omega =0$
instead of a maximum as expected for a FL. Simultaneously, Re$\Sigma ({\bf k}%
_F,\omega )$ has a positive slope near $\omega =0$. The dip in Im$\Sigma $
calculated within fRG is much more pronounced than in SOPT. Moreover, the
physical origin of these features is very different. The peculiarities of
the SOPT self-energy arise solely from the thermal excitation of electrons
near the vHs and although they exist at all $T>0,$ they lose their relevance
with decreasing T. Instead, the more pronounced
anomalies in the fRG self-energy {\it increase} in size with decreasing $T$ and
reflect the growing AF correlations and the concomitant
tendency towards pseudogap formation. Importantly, the single peak in
$A({\bf k}_F,\omega )=-$Im$G({\bf k}_F,\omega +{\rm i}0^{+})/\pi $ in Fig. 2 is
not a qp feature because the low energy structure of $\Sigma $ invalidates
the qp concept.

\begin{figure}[t!]
\psfig{file=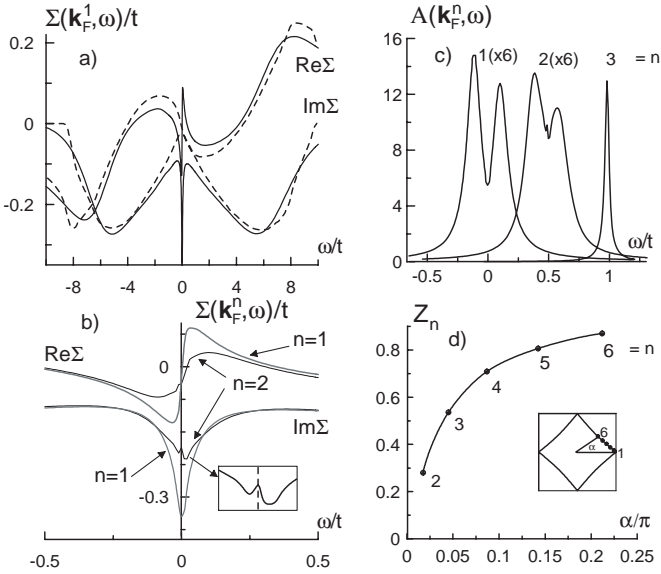,width=90mm,silent=} \vspace{2mm}
\caption{Same as Fig. 2 for $T=0.082t$ ($V_{max}=100t$). In addition, c) and
d) show the spectral functions and the qp weights for ${\bf k}_F^n$ in
different patches around the FS. The spectral functions are shifted by $%
(n-1)0.5t$ for a better view; for $n=1,2$ they are multiplied by 6.}
\label{fig:Fig22}
\end{figure}

In Fig. 3 we decrease the temperature to $T=0.082t$ which is closer to the
crossover temperature $T^{*}$ where the one-loop interaction vertices $V$
tend to diverge \cite{SalmHon} as a fingerprint of the nearby AF instability.
Although the low-temperature regime where the effective vertices are so large
($V_{\max }=100t$) is outside the validity region of the one-loop approach,
it proves most useful to clearly identify the structure of the spectral
function which originates from the above discussed form of the self-energy.
In the first patch a two-peak structure in $A({\bf k}_F,\omega)$ arises with
a local minimum at the Fermi energy. Outside the first patch, the real part
of the self-energy has a narrow region with negative slope near $\omega =0$
accompanied by a local maximum of Im$\Sigma $ (Fig. 3b), which leads to the
formation of a qp peak inside a pseudogap structure (Fig. 3c). This behavior
is in fact reminiscent of the finite temperature DMFT results in the metal
to insulator crossover region \cite{Bulla}. For larger $|\omega|$ the
behavior is
qualitatively similar in all patches. The qp peak quickly merges with the
two incoherent peaks of the pseudogap on approaching the BZ diagonal along
the FS. The qp weight around the FS (see Fig. 3d) gradually vanishes
with approaching the ($\pi ,0)$ point. For $t^{\prime }=0$ and the
corresponding vH band filling $n=1$, the qp peak is absent in all patches.
All spectral functions along the FS show in this case a two-peak pseudogap
structure at low temperatures similar to the TPSC results \cite{TPSC} and
the dynamical cluster approach \cite{DCA}.

In Figs. 4 a,b we show the self-energy for $U=2t$, $t^{\prime }=0.1t$ at
high and low temperatures, respectively, for $\mu =0.03t$ ($n=0.94$) when
the Fermi level is slightly above the vHs energy. At higher temperatures
only weak anomalies in $\Sigma ({\bf k}_F^1,\omega )$ are present near the
Fermi level. Upon lowering the temperature towards $T^{*}$ a clear maximum
of Im$\Sigma $ develops at the Fermi energy similar to that in Im$\Sigma
({\bf k}_F^n,\omega )$ for $n>1$ at vH band fillings. The real part of
self-energy has the proper negative slope for a FL in a narrow energy window
$|\omega |\lesssim \mu $ around the Fermi level. In the spectral function
(see Fig. 4c) we observe the split-off of two incoherent peaks near the qp
peak at the Fermi level. With increasing $\mu $ and the corresponding
increase of temperature which is kept above $T^{*}$, the qp peak gains weight
from the incoherent pseudogap peaks. For $\mu <0$ (fillings below vH band
filling) the value of $T^{*}$ rapidly drops with $|\mu |$ for finite
$t^{\prime },$ and we observe a quick crossover to a single-peak structure.

\begin{figure}[t!]
\psfig{file=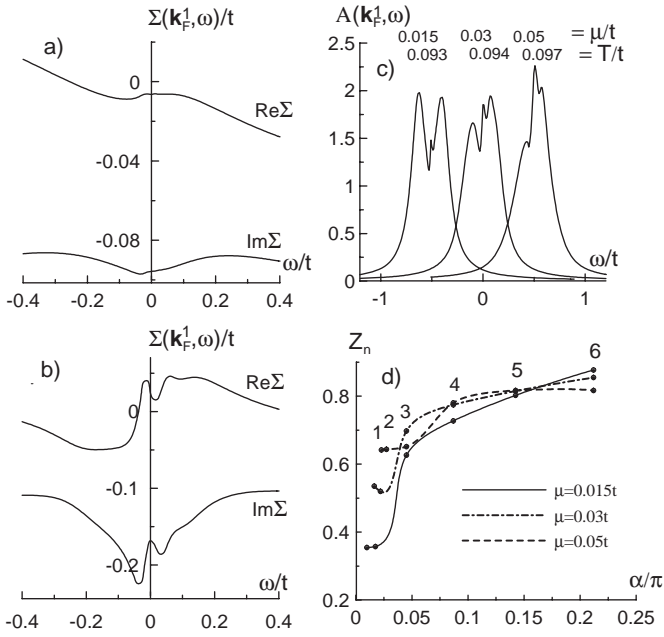,width=90mm,silent=} \vspace{2mm}
\caption{(a,b) Self-energy for a patch closest to the vHs at $U=2t$,
$t^{\prime }/t=0.1$, $\mu =0.03t$, $T=0.18t$ ($V_{max}=10t$; a) and $T=0.093t$
($V_{max}=100t$; b). (c,d) The evolution of the spectral function and the
quasiparticle weight for different chemical potentials, the temperatures
are chosen such that $V_{max}$=100t. The lines in (c) are shifted by
$0.5t$ for a better view. }
\label{fig:Fig3}
\end{figure}

Fig. 4d shows the anisotropic weight of the qp feature in the 3-peak spectral
functions of Fig. 4c for different chemical potentials $\mu >0$. Away from
the vH band filling the qp weight is finite in the first patch near the
$(\pi ,0)$ point, which is now below the Fermi level. Hot spots, i.e.
points on the FS which are connected by ${\bf Q}=(\pi ,\pi )$ and thus most
affected by the scattering from AF spin fluctuations, are identified as
local shallow minima in the angular dependence of the qp Z-factor along the
FS line. The central qp peak of the spectral function is suppressed
on parts of the FS between the near vicinity of hot spots and the FS points
closest to $(\pi ,0)$ and $(\pi ,0)$ and quickly gains spectral weight upon
moving towards the BZ diagonal.

In summary, we have investigated the self-energy on the real frequency axis
in the $2D$ $t$-$t^{\prime }$ Hubbard model
within a one-loop fRG analysis. For vH band fillings the self-energy has a
non-FL form at the FS points ($\pi ,0$) and ($0,\pi $) which are connected
by the AF wavevector ${\bf Q}$. Qps exist everywhere else on the FS but with
anisotropic spectral weight. The RG flow indicates that at low temperatures
the continuous decrease of the qp weight along the FS is accompanied by
the simultaneous growth of two additional incoherent peaks in the spectral
function. Away from vH fillings a qp peak at ($\pi ,0$) and ($0,\pi $)
emerges with small spectral weight.

These results provide a complimentary weak-coupling scenario for
pseudogap formation and offer
implications for the evolution of anisotropic spectral properties at finite
temperatures on the metallic side of the transition to an AF insulator.
The strong angular dependence of the qp properties in our fRG study
is reminiscent of the recent high-resolution ARPES studies on underdoped
cuprates \cite{ARPESAD}. Although the fRG method does not allow
to go below the crossover temperature $T^{*},$ and therefore to explore the
spectral functions at a fixed temperature in a wide range of dopings, we
expect further suppression of the weight of the central peak with decreasing
temperature and/or increasing interaction strength, and simultaneously
the increase of the doping range near half filling where the
pseudogap structures emerge.

The main change of the low-energy physics upon moving from the weak- to the
strong-coupling regime is expected in the redistribution of spectral weight
from the vicinity of the Fermi energy to the developing lower- and upper
Hubbard subbands. These subbands should form {\it in addition} to the
low-energy structures we have elucidated in our fRG analysis. This
conjecture finds indeed support from recent results of the dynamical cluster
approximation \cite{DCA} and cluster perturbation theory \cite{Senechal}.

The observed features are distinctly different from the previously found
low-temperature crossover to a non-FL form of the self-energy at half filling
\cite{FLEX,TPSC} as well as from the proposed hole pocket picture and the
partial destruction of the FS in strong coupling Hubbard or $t$-$J$ models
\cite{Lee}. The fRG results thereby offer a valid alternative for the
interpretation of experimental data from photoemission spectroscopy on
cuprates.

This work was supported by the Deutsche Forschungsgemeinschaft through SFB
484.

\end{document}